\documentclass[prb,preprint,showpacs]{revtex4}
\usepackage{graphicx}
\usepackage{epstopdf}
\usepackage{amsmath,amssymb}
\usepackage{bm}

\newcommand{\ignore}[1]{}
\newcommand{\mbold}[1]{\mbox{\boldmath $ #1 $}}

\newcommand{\beq}{\begin{equation}}
\newcommand{\eeq}{\end{equation}}

\begin{document}

\title{Termination of the Berezinskii-Kosterlitz-Thouless phase with a new critical universality in spin-crossover systems}
\author{Masamichi Nishino$^{1}$}
\email[Corresponding author: ]{nishino.masamichi@nims.go.jp} 
\author{Seiji Miyashita$^{2,3}$}
\affiliation{$^{1}${\it Computational Materials Science Center, National Institute}  
for Materials Science, Tsukuba, Ibaraki 305-0047, Japan \\
$^{2}${\it Department of Physics, Graduate School of Science,}
The University of Tokyo, Bunkyo-Ku, Tokyo, Japan \\
$^{3}${\it CREST, JST, 4-1-8 Honcho Kawaguchi, Saitama, 332-0012, Japan}
}
\date{\today}

\begin{abstract}
Two dimensional systems with U(1) symmetry exhibit a peculiar phase, i.e., the   Berezinskii-Kosterlitz-Thouless (BKT) phase. In particular situations, the BKT phase exists as an intermediate temperature phase. There have been scenarios for the phase transitions at the two endpoints of the intermediate BKT phase, i.e., the phase transition at the low-temperature endpoint is a BKT transition and that at the high-temperature endpoint is either a BKT transition or a first-order transition. The present study gives a novel scenario, i.e., a second-order transition with a new critical universality and a BKT transition. We found that this new phase transition is realized in spin-crossover systems on a triangular lattice with an antiferromagnetic short-range interaction. At the low-temperature transition the elastic interaction plays as a ferromagnetic infinite-range interaction and encourages the breaking of $Z_2$ symmetry between high-spin rich and low-spin rich states.
\end{abstract}


\maketitle

\section{Introduction}

In two-dimensional (2D) systems with U(1) symmetry, the Berezinskii-Kosterlitz-Thouless (BKT) transition~\cite{Berezinskii,Kosterlitz}, driven by topological defects,  appears, and it has been studied in many systems, such as superfluid helium films~\cite{Bishop}, superconducting films~\cite{Beasley}, two-dimensional melting~\cite{Halperin,Nelson,Young}, trapped atomic gases~\cite{Hadzibabic}, surface roughening~\cite{Chui}, etc. 

In particular situations, the BKT phase exists as an intermediate temperature phase. 
There have been scenarios for the phase transitions of the two endpoints of the intermediate BKT phase: (I) Jose, Kadanoff, Kirkpatrick and Nelson studied the BKT phase under discrete clockwise ($Z_p$ symmetry breaking)  perturbation 
by a Renormalization Group analysis~\cite{Jose}. They showed dual BKT transitions between a low-temperature ordered phase and a high-temperature disordered phase, where the $Z_p$ perturbation for $p \ge 5$ on the U(1) system becomes irrelevant for a temperature range below the critical temperature of the U(1) system. The critical exponents of the spin correlation function $\eta$, defined as $\langle \mbold{S}(\mbold{r}_i)\cdot\mbold{S}(\mbold{r}_j)\rangle\sim |\mbold{r}_i- \mbold{r}_j|^{-\eta}$, were estimated to be 1/4 and $4/p^2$ at the high and low end points of the intermediate phase, respectively. Following studies have obtained supporting results~\cite{Elitzur,Cardy}.  
The effect of $Z_p$ symmetry on phase transitions, especially $Z_6$ case, have been studied extensively. (II) In studies of 2D melting, the Kosterlitz-Thouless-Halperin-Nelson-Young theory~\cite{Halperin,Nelson,Young} presented the case of successive two kinds of BKT transitions with respect to the translational and orientational orders. 
(III) Recently, however,  Bernard and Krauth found another scenario with a first-order phase transition at the high-temperature endpoint of the BKT phase (hexatic phase) and a BKT transition at the low-temperature endpoint~\cite{Krauth}.

\begin{figure}[t]
\centerline{\includegraphics[clip,width=10cm]{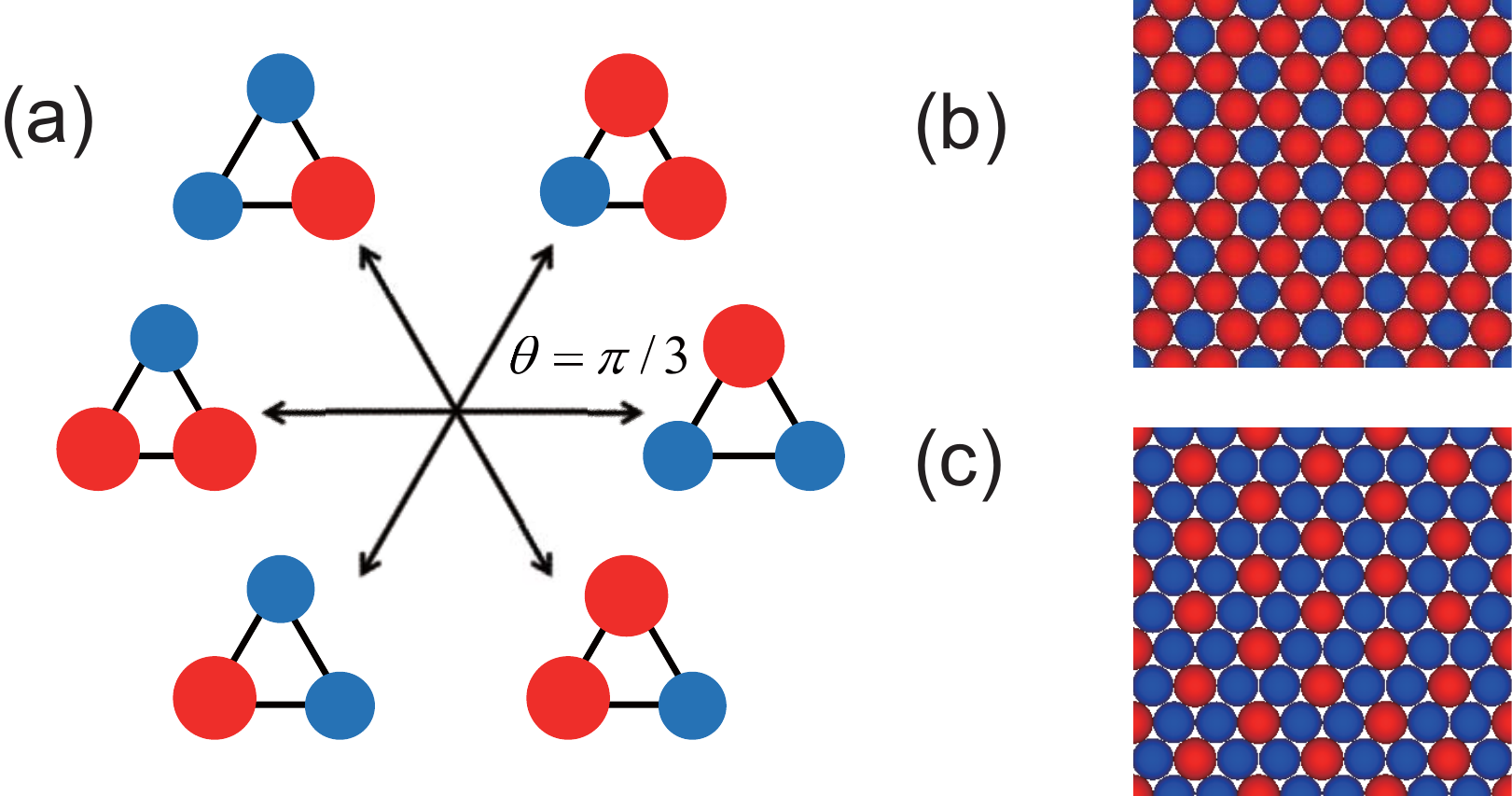}
}
\caption{ (color online) (a) Six plaquette states of the ferrimagnetic-like state.  A ground state configuration with (b) $\sigma_A=1$,$\sigma_B=1$, and $\sigma_C=-1$ and that with (c) $\sigma_A=-1$, $\sigma_B=-1$, and $\sigma_C=1$ are shown. Red and blue molecules denote high-spin and low-spin molecules, respectively. 
There is three fold degeneracy for configurations (b) and (c). }
\label{Fig_plaquette}
\end{figure}

The nature of the ordering process of the triangular Ising antiferromagnet 
with next-nearest-neighbor ferromagnetic interactions (TIAFF) has been studied in the picture of case (I)~\cite{Mekata,Schick,Landau,Takayama,Fujiki,Matsubara,Miyashita,Blote}. This model has six-fold degeneracy of the ground state and the six plaquette states $++-, -+-,\cdots$ (instead of red and blue molecules, $+$ and $-$ are allocated in Fig.~\ref{Fig_plaquette} (a)) are mapped to six state clock modes. 
An intermediate BKT phase has been observed between ferrimagnetic and 
disordered phases, while the model without the next nearest neighbor interaction does not exhibit any phase transition at finite temperatures~\cite{Wannier,Husimi,Houttappel}.

Recently, as a new aspect of phase transition, we have studied the effect of elastic interactions in the context of the spin-crossover (SC) material, in which the high spin (HS) and low spin (LS) states are expressed by an Ising spin. 
This material has attracted much attention in their potential applications to photo memory devices, etc. by making use of the nature of photoinduced phase transitions \cite{SC_book,SC_book2,Jakobi,Buron,Lorenc,Collet,Slimani,Enachescu,Nicolazzi}
 
Taking into account the molecular size difference between HS and LS molecules and the lattice deformation due to the difference, we have found that the elastic interaction plays an essential role in ordering process~\cite{Nishino1,Miya,Nakada,Nishino2}: the elastic interaction is relevant and its effective long-range nature is important in the ferromagnetic-like ordering~\cite{Miya,Nakada,Nishino2}, where volume fluctuation exists, while it is irrelevant in the antiferromagnetic-like ordering~\cite{Nishino2}, in which two ordered states have the same volume.

Here we study the effect of the elastic interaction on the TIAFF model, which is a prototype modeling for triangular SC materials with frustration. 
In this work we present a new scenario of criticality for the endpoints of the BKT phase: a second-order transition with a new universality class at the low-temperature endpoint and a BKT transition at the high-temperature endpoint. 
We show that this novel phenomenon is induced by the synergetic effect of frustration and the elastic interaction, which we find equivalent to an effective long-range ferromagnetic interaction.

The rest of the paper is organized as follows. 
In Sec.~\ref{sec_model} the SC model including the elastic interaction is presented and the method to analyze the critical properties is given.
In Sec.~\ref{sec_6fold} we study the ground-state properties of the model. 
In Sec.~\ref{sec_BKT_phase} we show novel critical properties of the intermediate temperature BKT phase due to the elastic effect. 
In Sec.~\ref{sec_LRnature} we show that the elastic interaction is expressed by an infinite long-range interaction and this long-range interaction affects the 
criticality at the low-temperature end point. 
Section~\ref{sec_exponents} is devoted to the analyses of the critical exponents at the low-temperature end point. 
In Sec.~\ref{summary} we give a summary.

\section{Model and method}
\label{sec_model}

We adopt the following Hamiltonian in the triangular lattice, which consists of the short-range interactions (TIAFF model) ${\cal H}_{\rm I}$  and elastic interactions ${\cal H}_{\rm el}$: 
\begin{equation}
{\cal H}=  {\cal H}_{\rm I}+{\cal H}_{\rm el},
\label{Ham}
\end{equation}
where
\begin{eqnarray}
{\cal H}_{\rm I}= &&- J_1\sum_{\langle i ,j \rangle }\sigma_{i} \sigma_{j}
- J_2\sum_{\langle \langle i ,k \rangle \rangle}\sigma_{i} \sigma_{k}
\label{Ham_I}
\;\;\;  \\
{\rm and} \;\;\;\;\;\;\;\;  {\cal H}_{\rm el}= && {k \over 2}\sum _{\langle i,j \rangle}[r_{i,j}-(R(\sigma_i)+R(\sigma_j))]^2. 
\label{Ham_el}
\end{eqnarray}
 Here $r_{i,j}=|\vec{x}_i-\vec{x}_j |$ is the distance between the $i$th and $j$th molecules, where $\vec{x}_i$ denotes the position of the $i$th molecule. 
Each molecule $i$ has the LS state, denoted by the pseudo spin $\sigma_i=-1$, or the HS state, $\sigma_i=1$. 
The HS molecule has a larger radius $R_{\rm H}$ than that of the LS molecule $R_{\rm L}$, and we express $R(-1)=R_{\rm L}$ and $R(1)=R_{\rm H}$. 
Parameter $k$ is the elastic constant, $J_1(<0)$ is the 
antiferromagnetic Ising interaction between nearest neighbors $\langle i ,j \rangle$, and $J_2(>0)$ is the ferromagnetic Ising interaction between next nearest neighbors $\langle \langle i ,k \rangle \rangle$. 
In the SC system, the energy difference of the HS and LS states, denoted by $D$,  and the ratio of degeneracy $g$ give an effective field~\cite{SC_book2} but here we focus on the critical phenomena along the coexistence line and set the field to be zero. 
Here we adopt the parameters as $k = 40$, $R_{\rm L}=1$, $J_1=-0.1$, $J_2=0.02$.

As long as the ratio $R_{\rm H}/R_{\rm L}$ is not far from 1, 
the system has 6-fold degenerate ferrimagnetic-like ground states ($Z_6$ symmetry) (see Figs.~\ref{Fig_plaquette} (a)-(c) and Sec.~\ref{sec_6fold}). 
Here the six states are characterized by three sublattice states (three sublattices are called sublattice A, B, and C) as in Fig.~\ref{Fig_plaquette} (a). We find a characteristic of 
volume difference between three HS-rich and three LS-rich states. 
If the ratio $R_{\rm H}/R_{\rm L}=1$, the elastic interaction term (\ref{Ham_el}) does not contribute to the ordering processes in this study, and we have confirmed that the model shows the same critical phenomena as the TIAFF model (\ref{Ham_I}) (not shown).

To observe the ordering process of the model, we introduce a vector which quantifies the three-sublattice states shown in Fig.~\ref{Fig_plaquette} (a): 
\begin{equation}
\vec{v}_m \equiv (\cos \theta_m, \sin \theta_m) \;\; {\rm for} \;\; \theta_m={2\pi \over 6} k,\; k=0,1,\cdots, 5 
\end{equation}
at the $m$th triangular plaquette. 
The vectors at the plaquettes align in one of six direction in the ground state and the model can be regarded as a six-state clock model. 
Now we introduce an order parameter which characterizes the degree of the order as
\begin{equation}
M^2= \Big \{\sum_{m=1}^{ N_{\rm p}} \cos \theta_m \Big \}^2+ \Big \{\sum_{m=1}^{ N_{\rm p}} \sin \theta_m \Big \}^2. 
\end{equation}
Here $N_{\rm p}$ is the number of the triangular plaquettes ($N_{\rm p}=N/3=
L^2/3$), where $N$ is the number of molecules and $L$ is the linear dimension of the system. 
We also observe the quantity  
\begin{equation}
m=\sum_i^{N} \sigma_i, 
\end{equation} 
which relates to the HS fraction and represents volume fluctuation.

In the present work, we applied a Monte Carlo method to obtain physical quantities with $NPT$ (isothermal-isobaric) ensemble, in which the pressure is set to $P=0$ with periodic boundary conditions~\cite{Nishino2}.

\section{Six-fold degeneracy of the ground state}
\label{sec_6fold} 

We obtained the ferrimagnetic-like configurations of HS and LS molecules 
as shown in Figs.~\ref{Fig_plaquette} (b) and (c) at low temperatures in the simulations of the elastic model (\ref{Ham}). 
There is three fold degeneracy between the states for  
the group A: $\theta=\pi/3$, $\theta=\pi$, and $\theta=5\pi/3$ (Fig.~\ref{Fig_plaquette} (a) ) and also between the states for the group B: $\theta=0$, $\theta=2\pi/3$, and $\theta=4\pi/3$. 

First, we examine the ground-state energy of the elastic model (\ref{Ham}) for groups A and B for the case $R_{\rm H}/R_{\rm L}\neq 1$. 
If $R_{\rm H}/R_{\rm L} = 1$, the ground-state energy of the elastic model (\ref{Ham}) is the minimum energy of the TIAFF model (\ref{Ham_I}) and the ferrimagnetic-like state is the ground state. 
If $R_{\rm H}/R_{\rm L}\neq 1$, the elastic energy contributes. 
The elastic energy for a unit triangular plaquette is     
\begin{eqnarray}
U= && \frac{k}{2} \Big[r_{\rm AB} - (R_{\rm A} + R_{\rm B}) \Big]^2 + \frac{k}{2} \Big[r_{\rm BC} - (R_{\rm B} + R_{\rm C}) \Big]^2  +  \frac{k}{2} \Big[r_{\rm CA} - (R_{\rm C} + R_{\rm A}) \Big]^2, 
\label{ene_triangle}
\end{eqnarray}
where $R_{\rm A}$, $R_{\rm B}$, and $R_{\rm C}$ are the radii of the molecules of sublattices A, B, and C, respectively, and $r_{\rm AB}$ denotes the distance between the center of the molecule of sublattice A and that of sublattice B, and 
$r_{\rm BC}$ and $r_{\rm CA}$ are defined in the same manner. 
It should be noted that when the ratio $R_{\rm H}/R_{\rm L}$ is extremely large, the ground state is the complete HS or LS state (uniform configuration), which we do not treat in this study. 

In the ground state, the unit triangle has LS, HS, and HS molecules in group A and HS, LS, and LS molecules in group B. As we see in Fig.~\ref{Fig_plaquette} (b) and (c), the configurations have $C_6$ symmetry concerning the LS and HS molecules, respectively, and the triangle of the unit cell is an equilateral triangle in both cases. 
We define 2$x$ = $r_{\rm AB}(=r_{\rm BC}=r_{\rm CA})$  as the length of a side of the triangle. 

In group A, the radii of the molecules are $R_{\rm L}$, $R_{\rm H}$, and $R_{\rm H}$ and the energy is given by 
\begin{eqnarray}
U= && \frac{k}{2} \Big[2x - (R_{\rm L} + R_{\rm H}) \Big]^2 + \frac{k}{2} \Big[2x - 2R_{\rm H} \Big]^2  +  \frac{k}{2} \Big[2x - (R_{\rm H} + R_{\rm L}) \Big]^2 \nonumber  \\      
 = && \frac{k}{2} \Big[ 3 \Big(2x-2R_{\rm H}+ \frac{2}{3} \delta \Big)^2 + \frac{2}{3} \delta^2  \Big]  \ge \frac{1}{3}k \delta^2, 
\end{eqnarray}
where $\delta \equiv R_{\rm H} -R_{\rm L}$. 
Thus the minimum energy is $\frac{1}{3}k \delta^2$ with $x=R_{\rm H}- \frac{1}{3} \delta$. 
On the other hand, in group B, the radii of the molecules are $R_{\rm H}$, $R_{\rm L}$, and $R_{\rm L}$ and the energy is given by 
\begin{eqnarray}
U= && \frac{k}{2} \Big[2x - (R_{\rm H} + R_{\rm L}) \Big]^2 + \frac{k}{2} \Big[2x - 2R_{\rm L} \Big]^2  +  \frac{k}{2} \Big[2x - (R_{\rm L} + R_{\rm H}) \Big]^2  \nonumber  \\
 = && \frac{k}{2} \Big[ 3 \Big(2x-2R_{\rm L}- \frac{2}{3} \delta)^2 + \frac{2}{3} \delta^2  \Big]  \ge \frac{1}{3}k \delta^2, 
\end{eqnarray}
where the minimum energy is $\frac{1}{3}k \delta^2$ with $x=R_{\rm L}+ \frac{1}{3}\delta$. 
Thus we find the same minimum energy $\frac{1}{3}k (R_{\rm H} -R_{\rm L})^2$ in both groups, where $2x=2R_{\rm H}- \frac{2}{3} (R_{\rm H} -R_{\rm L})$ for group A and $2x=2R_{\rm L}+ \frac{2}{3} (R_{\rm H} -R_{\rm L})$ for group B. 
In each case the equilibrium of the six forces acting on each molecule is easily confirmed.

Next we study the ground state entropy of the elastic Hamiltonian for groups A and B in the harmonic approximation of the deviations from the ground state configuration. 
The elastic Hamiltonian ${\cal H}_{\rm el}$ is expressed as
\begin{eqnarray}
{\cal H}_{\rm el}=U_0 + {}^t\! \vec{x} A \vec{x},
\end{eqnarray}
where $U_0=\frac{1}{3}k (R_{\rm H} -R_{\rm L})^2 N_{\rm p}$ is the ground-state energy, $\vec{x}$ is defined as $\vec{x}\equiv (\delta \vec{x}_1,\delta \vec{x}_2,\cdots,\delta \vec{x}_N)$, and the matrix $A$ is a $2N \times 2N$ matrix which gives 2nd-order expansion coefficients.

Since the partition function for the Hamiltonian at $\beta=\frac{1}{k_{\rm B} T}$ ($k_{\rm B}=1$ is set) is given by 
\begin{eqnarray}
Z=\int_{-\infty}^{\infty} d\vec{x} \exp(-\beta (U_0+ {}^t\! \vec{x} A \vec{x}   ))=e^{-\beta U_0} \frac{\pi^{\frac{N}{2}}}
{\beta^{\frac{N}{2}} (\det A)^{\frac{1}{2}} } ,
\end{eqnarray}
the free energy is given by 
\begin{eqnarray}
F=&&-\frac{1}{\beta} \ln Z   \nonumber      \\
  =&& U_0-\frac{1}{\beta} \Big( \frac{N}{2} \ln \big(\frac{\pi}{\beta}\big)-\frac{1}{2} \ln (\det A) \Big). 
\end{eqnarray}
Thus the entropy is 
\begin{eqnarray}
  S=&& \frac{N}{2} \ln \big(\frac{\pi}{\beta}\big)-\frac{1}{2} \ln (\det A)  \nonumber  \\
   =&&  \frac{N}{2} \ln \big(\frac{\pi}{\beta}\big)-\frac{1}{2} \sum_{\vec{k}} \ln (\det A(\vec{k})). 
\end{eqnarray}

Making use of the periodicity of the lattice, we estimated $S$ 
in the $\vec{k}$ space, where the sum runs over the first Brillouin zone. Applying Fourier transformation for each sublattice~\cite{Miya2}, we have a $6 \times 6$ matrix for $A(\vec{k})$. 
We computed numerically the entropies of groups A and B ($S_{\rm A}$ and $S_{\rm B}$) and found that $S_{\rm B}$ is larger than $S_{\rm A}$. Thus the ground state entropy depends on the configuration of group A or B. 
In the present work, however, we performed simulations of the elastic model, decreasing the temperature slowly to obtain the ferrimagnetic-like ordered state, and we observed almost equal frequency for configurations of A and B. This indicates that at the finite temperatures, all the six states are nearly degenerate. 
Here the system does not choose the B type ferrimagnetic-like state by the order by disorder mechanism. The dependence of the frequency for configurations A and B on the cooling procedure will be reported elsewhere. 
In Secs~\ref{sec_LRnature} and \ref{sec_exponents}, we also find that the infinite-range model (\ref{Ham_prime}), in which the six states are definitely degenerate, exhibits the same type critical behavior. Thus we concluded that the six states of the elastic model can be regarded to be degenerate at the lower critical point ($T_{\rm c2}$). 
 Moreover, the coincidence of the critical behavior in both models indicates a new universal criticality, as we show in the following sections.

\section{Effect of the Elastic interaction on the BKT phase}
\label{sec_BKT_phase}

\subsection{Intermediate temperature BKT phase}

As mentioned in the introduction, the TIAFF model shows dual BKT transitions, and there exists an intermediate BKT phase ($T_{\rm c2} \le T \le T_{\rm c1}$) between the ferrimagnetic and paramagnetic phases. The spin correlation function decays in a power law with temperature-dependent exponent $\eta(T)$~\cite{Jose,Cardy}:  $\eta(T_{\rm c2})=1/9 \le \eta(T) \le \eta(T_{\rm c1})=1/4$.

In the Monte Carlo simulation, $\eta$ is estimated from the following quantity, 
\begin{equation}
a(L,L')=\frac{ \ln \frac{\langle M_L^2 \rangle}{ L^2 } /\ln \frac{\langle M_{L'}^2 \rangle}{ L'^2 }      }{  \ln(L/L')}, 
\label{aL}
\end{equation}
where $L$ and $L'$ denote the linear dimension of the lattice and 
$\langle X \rangle$ is the thermal average of $X$. 
This quantity gives a crossing point which indicates the critical temperature, and the value at this point gives $2-\eta$. 

\begin{figure}[t]
  \begin{center}
     \includegraphics[width=53mm]{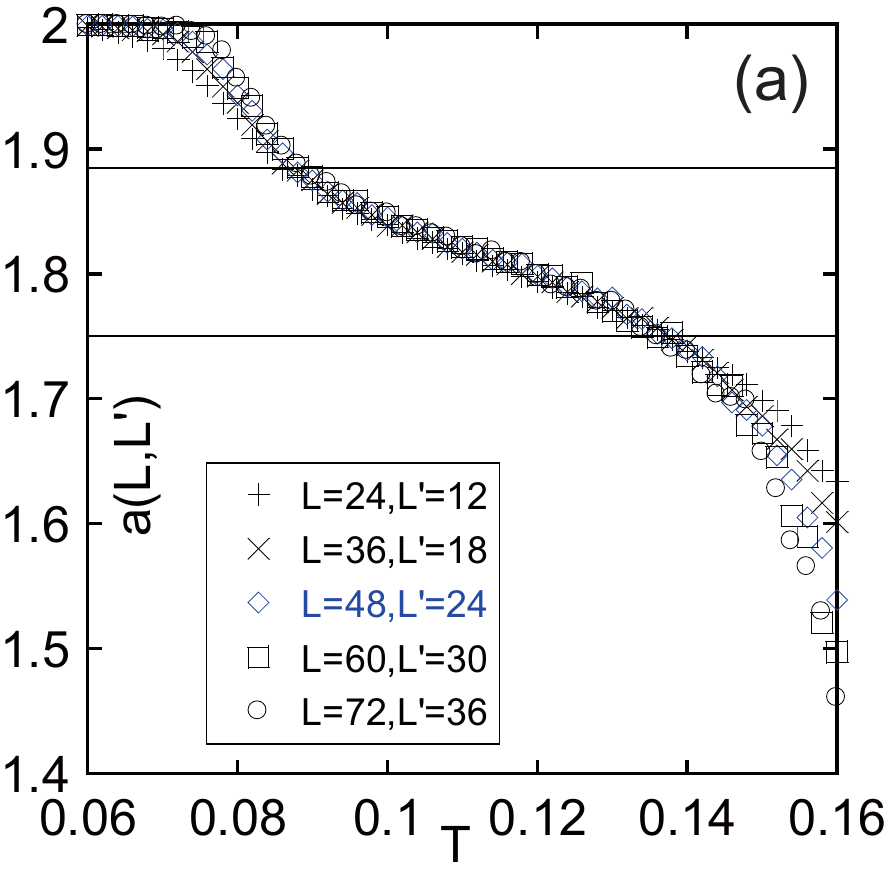}
     \includegraphics[width=53mm]{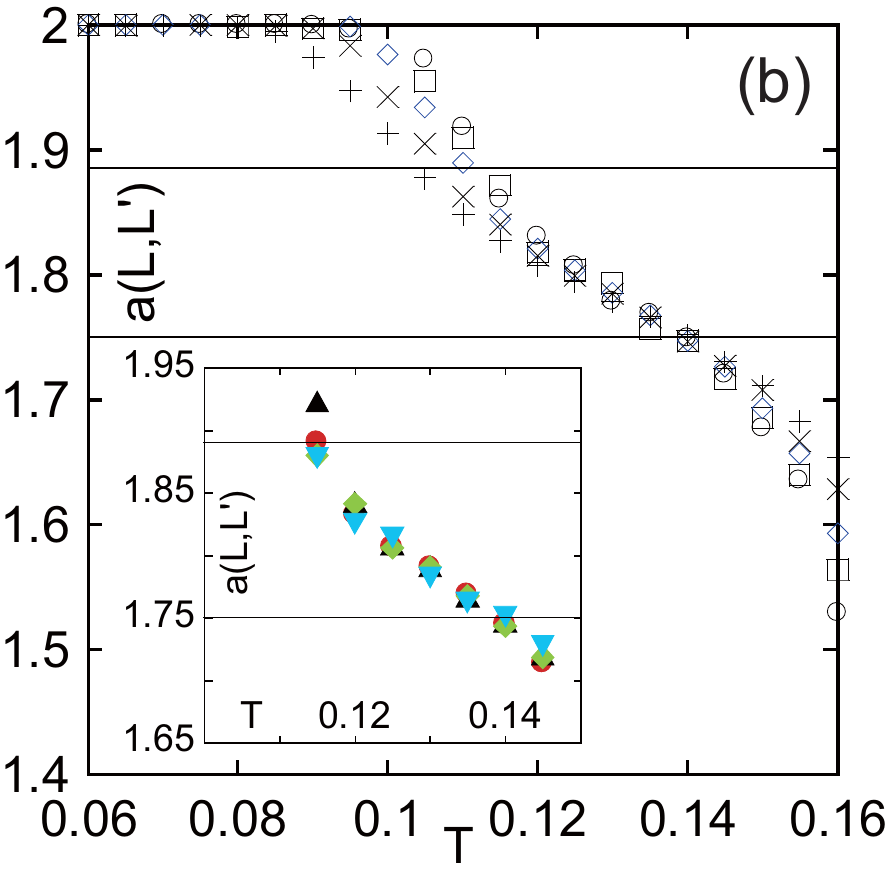}
\hspace{1.0cm}
     \includegraphics[width=55mm]{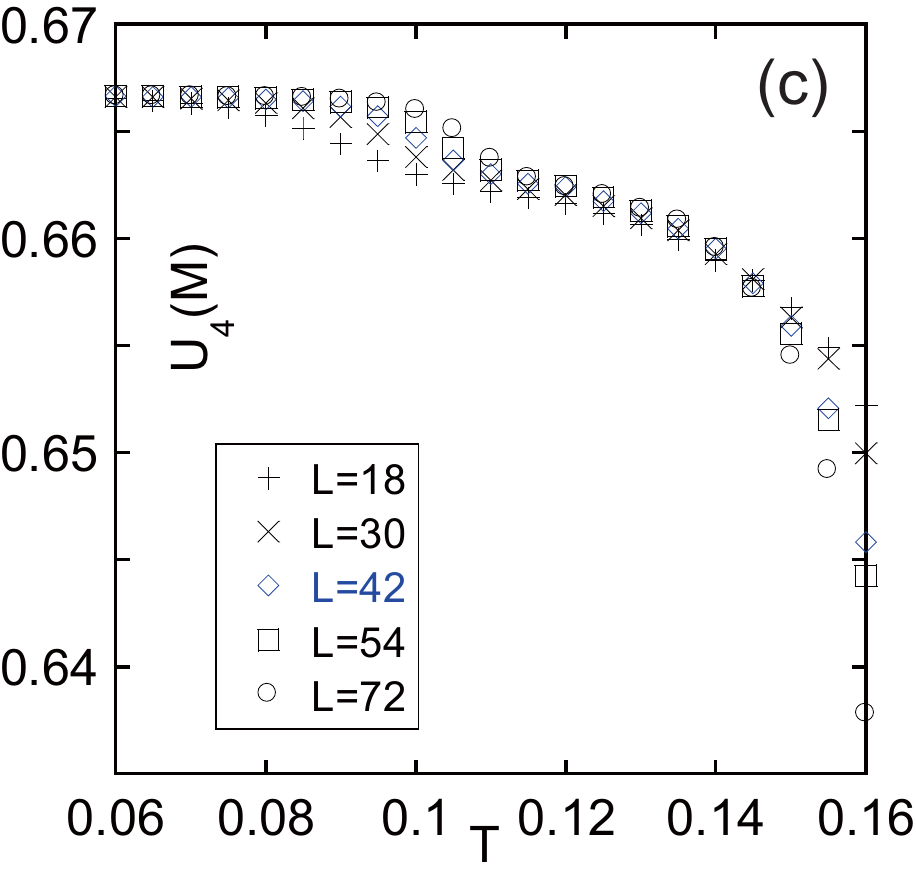}
     \includegraphics[width=53mm]{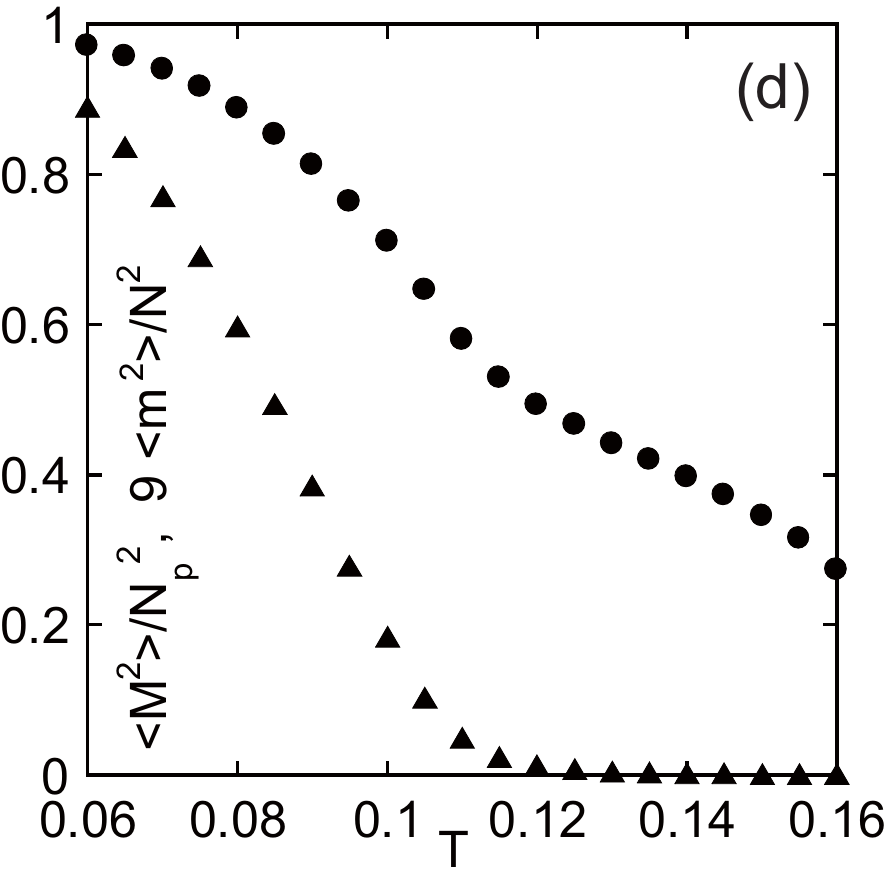}
  \end{center}
\caption{ (color online) 
(a) $a(L,L')$ as a function of $T$ of the TIAFF model (2), which corresponds to the SC model (1) with $R_{\rm H}/R_{\rm L}=1$. The upper and lower horizontal lines correspond to the locations of $\eta=1/9$ and $\eta=1/4$, respectively. 
 (b) $a(L,L')$ as a function of $T$ for $R_{\rm H}/R_{\rm L}=1.03$. 
The symbols for the size dependence are the same as (a). 
The inset shows $a(L,L')$ for $(L,L')=(84,42)$ (down-pointing triangles), $(96,48)$ (diamonds), $(108,54)$ (triangles), and $(120,60)$ (circles). 
(c) $U_4(M)$ as a function of $T$ for $R_{\rm H}/R_{\rm L}=1.03$.
(d) $\langle M^2 \rangle/{N}_{\rm p}^2$ ($\bullet$) and $9 \times \langle m^2 \rangle/N^2$ ($\blacktriangle$) as a function of $T$ for $R_{\rm H}/R_{\rm L}=1.03$. $L=72$.}
\label{Fig_eta}
\end{figure}

We give $a(L,L')$ for the TIAFF model (\ref{Ham_I}) in Fig.~\ref{Fig_eta} (a) as a function of the temperature $T$ for several systems sizes. 
The data overlap well in an intermediate-temperature region (between $T \simeq 0.138$ and $T \simeq 0.088$) 
and the value of $\eta(T)$ changes consistently with the above-mentioned property. Here we used 400,000 Monte Carlo steps (MCS) for the equilibration and the following 400,000 MCS to obtain $a(L,L')$. 
We confirmed that the elastic model (\ref{Ham}) gives the same results if we set $R_{\rm H}/R_{\rm L}=1$  (not shown). 

Next we study the effect of the elastic interaction by setting $R_{\rm H}/R_{\rm L}>1$. 
Here we adopt $R_{\rm H}/R_{\rm L}=1.03$ as a representative of the weak elastic interaction. It should be noted that we found that the strong elastic interaction, e.g., $R_{\rm H}/R_{\rm L} \sim 1.1$, causes a first-order phase transition between the disordered and ferrimagnetic-like phases and the BKT phase does not appear. 
We depict $a(L,L')$ in Fig.~\ref{Fig_eta} (b). 
Here we used 1,000,000$\sim$2,000,000 MCS for the equilibration and the following 1,000,000$\sim$8,000,000 MCS for observation. 
An overlap of $a(L,L')$ is found between $T \simeq 0.120$ and $T \simeq 0.140$ in Fig.~\ref{Fig_eta} (b), and the overlap is clear in larger system sizes as shown in the inset of Fig.~\ref{Fig_eta} (b). 
Thus, we conclude that a part of BKT phase remains. 
In contrast to the case $ R_{\rm H}/R_{\rm L}=1$, however, the overlap terminates before $\eta(T)$ reaches 1/9. 

 In Fig.~\ref{Fig_eta} (c), we also plot the Binder cumulant of the order parameter $M$, 
\begin{equation}
U_4(M)=1-\frac{ \langle M^4 \rangle   }{ 3 \langle M^2 \rangle^2   }. 
\end{equation}
Here $U_4(M)$ for different sizes also overlap in the same temperature region 
as $a(L,L')$. 
This observation supports the existence of the BKT phase. 

It should be noted that, because of the nature of the BKT phase transition, 
precise determination of the termination point is difficult from Figs.~\ref{Fig_eta} (b) and (c), and also from $\langle M^2 \rangle/N^2_{\rm p}$ vs $T$ curves (see Fig.~\ref{Fig_eta} (d)). 

\subsection{Uniform Magnetization}
\label{sec_Unimag}

\begin{figure}[t]
  \begin{center}
     \includegraphics[width=53mm]{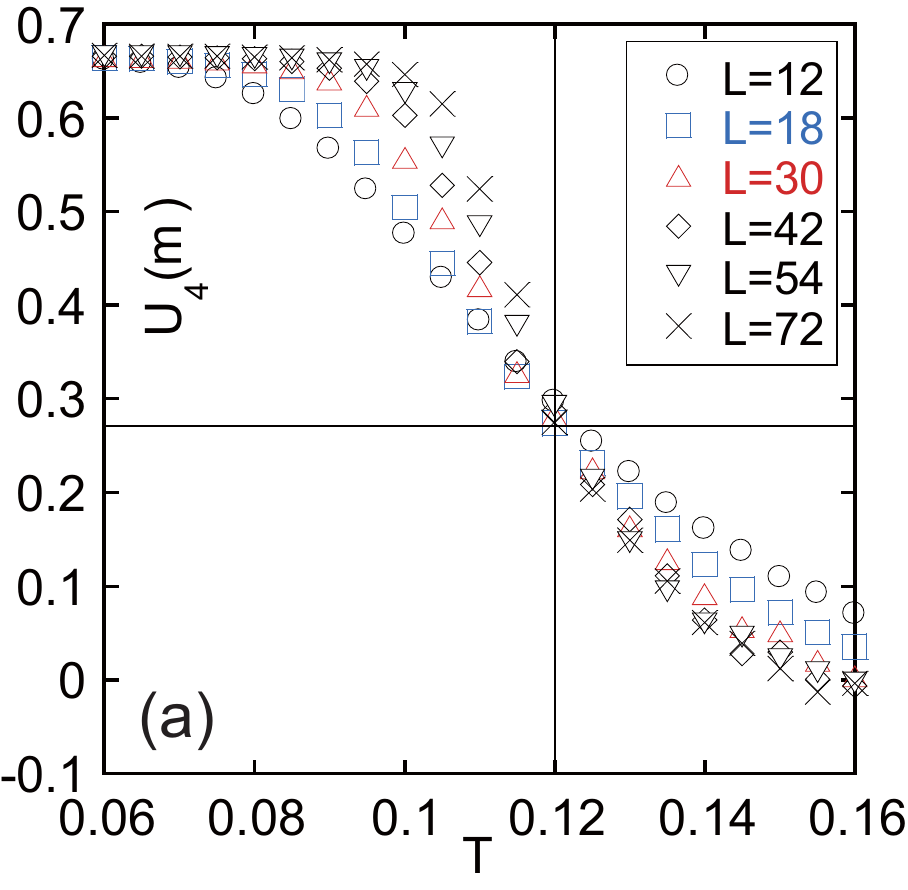}
     \includegraphics[width=53mm]{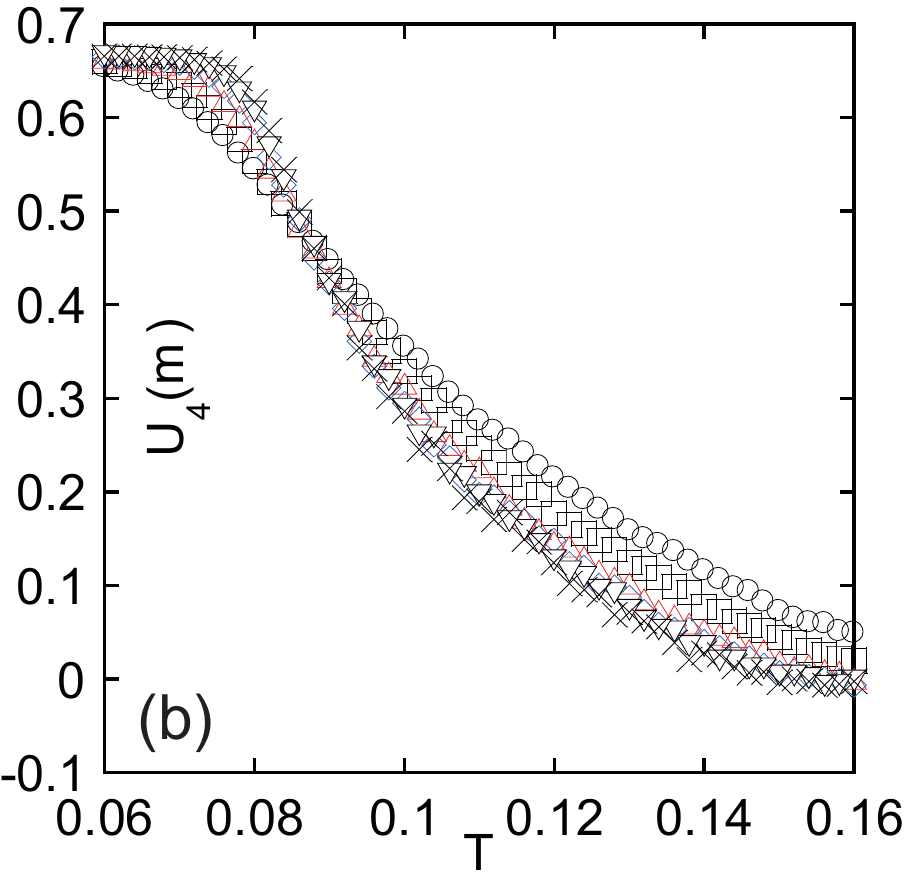}
  \end{center}
\caption{ (color online)  (a) $U_4(m)$ as a function of $T$ for $R_{\rm H}/R_{\rm L}=1.03$. The horizontal line corresponds to $U_4(m)=0.271$, which is the value of the mean-field theory. (b) $U_4(m)$ as a function of $T$ for the TIAFF model (\ref {Ham_I}), which corresponds to the case of $R_{\rm H}/R_{\rm L}=1$.  The correspondence between the symbol and the system size is the same as (a). 
 }
\label{Fig_mag}
\end{figure}

In the low-temperature phase, the ferrimagnetic-like state is realized, and the uniform magnetization $m$ appears (Fig.~\ref{Fig_eta} (d)). 
We make use of this fact, and then determine the critical point from the Binder plot of $m$ (Fig.~\ref{Fig_mag} (a)): 
$U_4(m)=1-\langle m^4 \rangle/3 \langle m^2 \rangle^2$. 
The data show a clear single crossing at $T \simeq 0.120$ and $U_4(m) \simeq 0.27$. 
For comparison, we give the Binder plot of the TIAFF model (\ref{Ham_I}) in Fig.~\ref{Fig_mag} (b). 
We find a single crossing, but $U_4(m)\simeq 0.46$ at $T_{\rm c2}$, which is different from that in Fig.~\ref{Fig_mag} (a).
We confirmed that the value at a single crossing $U_4(m) \simeq 0.46$ for the case $R_{\rm H}/R_{\rm L}=1$. 
The difference of $U_4(m)$ indicates that the elastic interaction ($R_{\rm H}/R_{\rm L} \neq 1$) causes a qualitative change of the nature of the phase transition at $T_{\rm c2}$. 

Now we are interested in the critical behavior of $m$.
In Fig.~\ref{Fig_mag} (a), we found that $U_4(m)$ at the crossing point is close to that of the ferromagnetic mean-field (MF) model. The appearance of the spontaneous magnetization of the MF theory is given by $m \sim |T-T_{\rm c}|^{\beta}$ with $\beta=1/2$. However, in Fig.~\ref{Fig_eta} (d), we find $m^2$ shows a convex shape, 
but not a linear shape for the case $\beta=1/2$. This means that critical exponent $\beta > 1/2$. 
This fact indicates that the critical nature in the present case is not explained by the MF theory to simple ferromagnetic systems, and presents a new type of criticality (see Sec.~\ref{sec_exponents}).

\section{Long-range nature of the Elastic interaction}
\label{sec_LRnature}

\begin{figure}[t]
  \begin{center}
      \includegraphics[width=53mm]{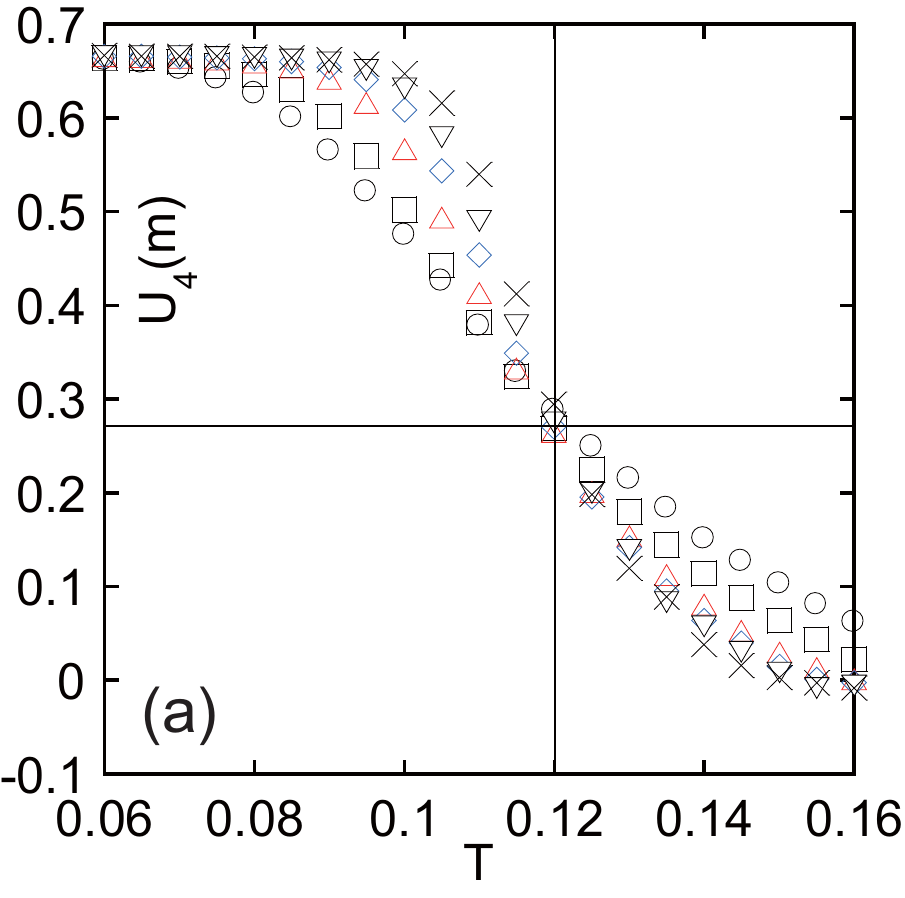}
      \includegraphics[width=54mm]{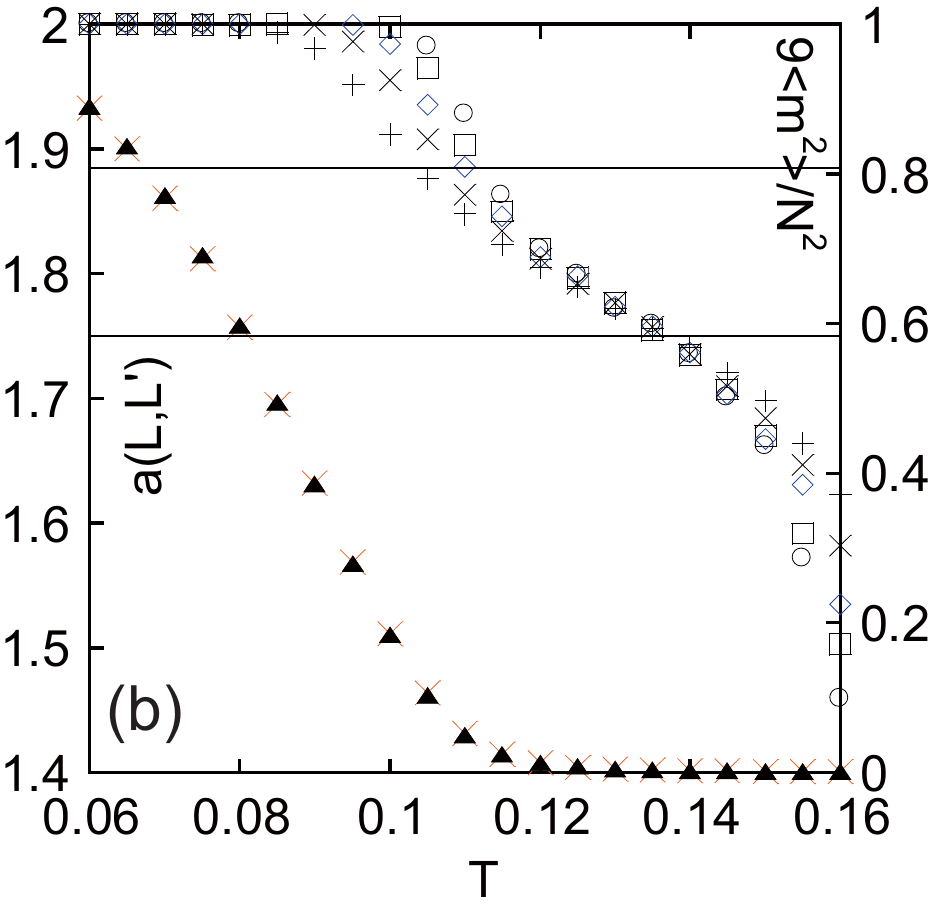}
  \end{center}
\caption{ (color online) (a) $U_4(m)$ as a function of $T$ for 
$J_{\rm inf}=0.042$ in the LRI model (\ref{Ham_prime}). 
The horizontal line corresponds to $U_4(m)=0.271$, which is the value of the mean-field theory. The symbols for the size dependence are the same as Fig.~\ref{Fig_mag} (a). 
(b) $a(L,L')$ as a function of $T$ for $J_{\rm inf}=0.042$ in the LRI model (\ref{Ham_prime}). The symbols for the size dependence are the same as Fig.~\ref{Fig_eta} (b). 
 $9 \times \langle m^2 \rangle/N^2$ (red crosses) for $L=72$ as a function of $T$ for the LRI model (\ref{Ham_prime}). Black triangles are the same as in Fig.~\ref{Fig_eta} (d). Red crosses and black triangles well overlap. 
}
\label{Fig_mag_RLmodel}
\end{figure}

We consider that some long-range nature of the elastic interaction is a key for the realization of this criticality. 
To clarify this point, 
we study the following long-range interaction model (LRI model):  
\begin{eqnarray}
{\cal H'}={\cal H}_I + {\cal H}_{\rm inf}, 
\label{Ham_prime}
\end{eqnarray}
where 
\begin{eqnarray}
{\cal H}_{\rm inf}= &&- \frac{J_{\rm inf}}{N} \sum_{i < j}\sigma_{i} \sigma_{j}. 
\label{Ham_inf}
\end{eqnarray}
Here the sum in Eq.~(\ref{Ham_inf}) runs over all the pairs. 
We set $J_{\rm inf}=0.042$ (ferromagnetic) to adjust the critical temperature ($T_{\rm c2}$) close to $T\simeq 0.12$. 
We depict $U_4(m)$ in Fig.~\ref{Fig_mag_RLmodel} (a) for the LRI model (\ref{Ham_prime}), where we find surprisingly a very similar temperature dependence to that in Fig.~\ref{Fig_mag} (a). 

In Fig.~\ref{Fig_mag_RLmodel} (b) we plot $a(L,L')$ of this model, which is also very similar to that of the elastic model (\ref{Ham}). The temperature dependence of $\langle m^2 \rangle$ for $L=72$ is also plotted (red crosses). It shows an excellent agreement with that obtained in Fig.~\ref{Fig_eta} (d) (black triangles). Furthermore, we find that the two models have the same critical exponents as we show in the next section. 
Thus, we conclude that
this LRI model (\ref{Ham_prime}) is an effective model for the elastic model (\ref{Ham}), and that the elastic interaction plays a role of the ferromagnetic infinite-range interaction. This fact indicates that the competition between the short-range frustrated antiferromagnetic interaction and the long-range ferromagnetic interaction is the key mechanism for the present new critical behavior.

\section{Critical exponents at the low-temperature end point}
\label{sec_exponents}

\subsection{Critical exponents for the elastic model}

\begin{figure}[t]
  \begin{center}
     \includegraphics[width=60mm]{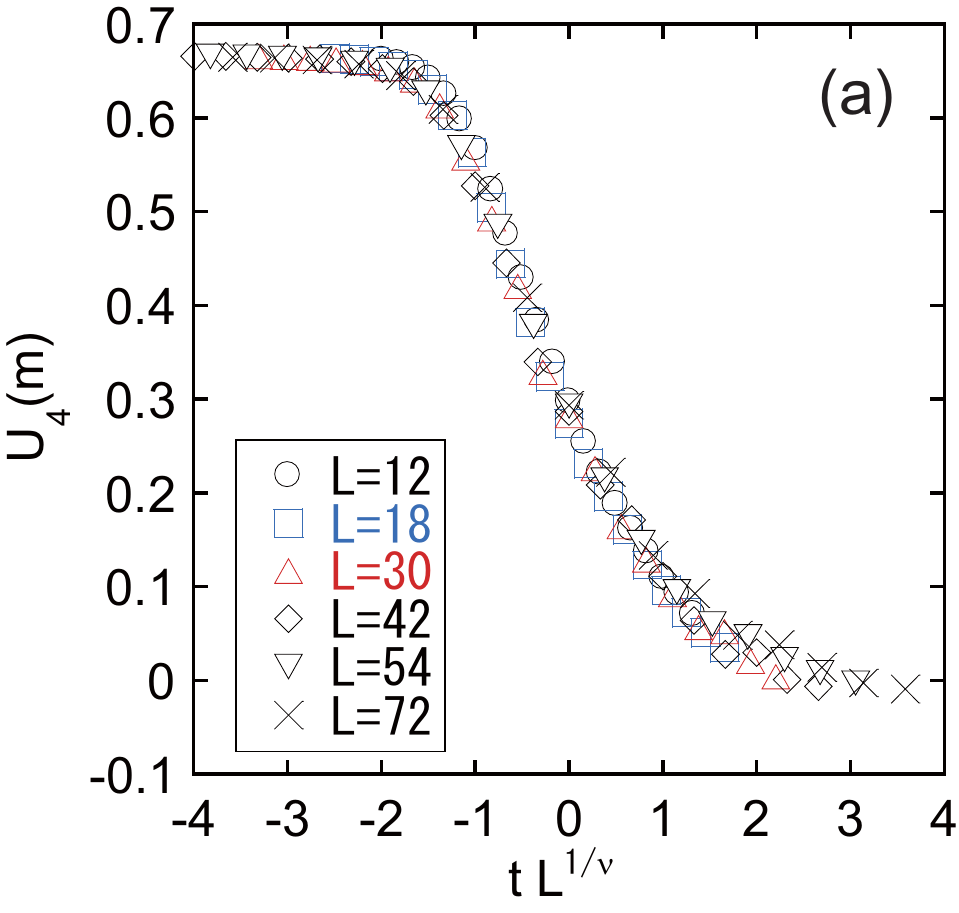}
     \includegraphics[width=62mm]{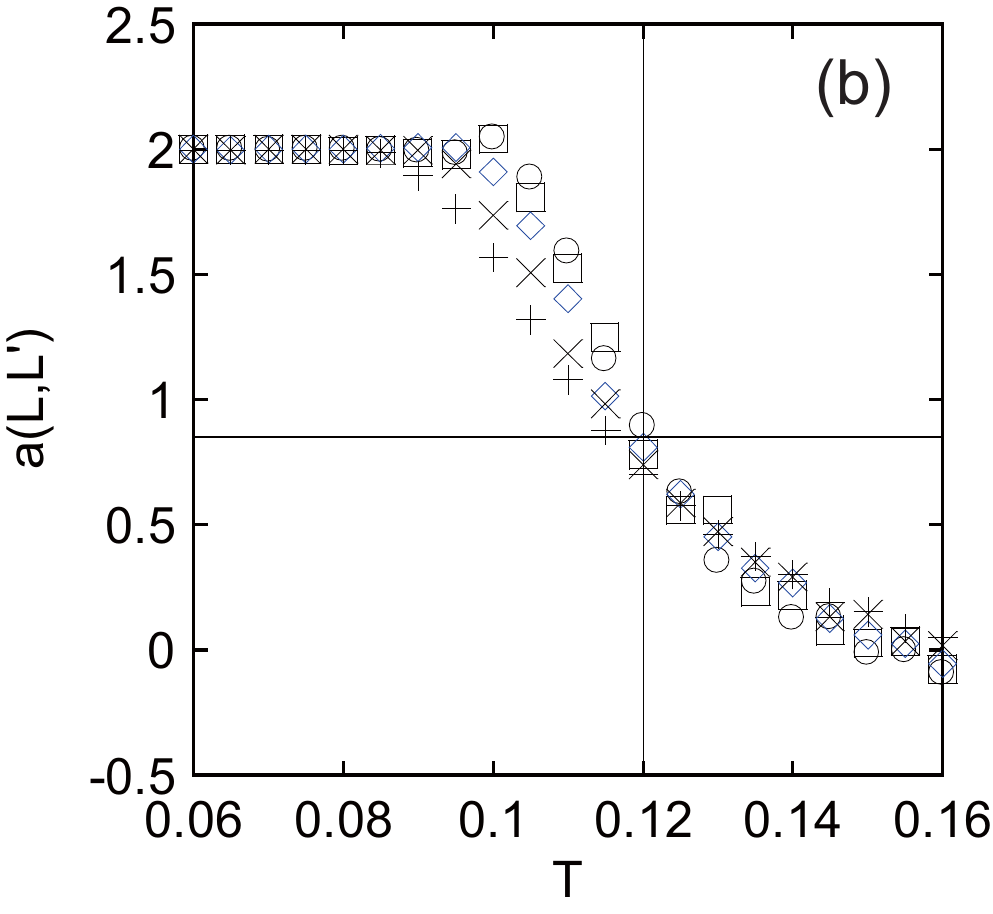}
     \includegraphics[width=60mm]{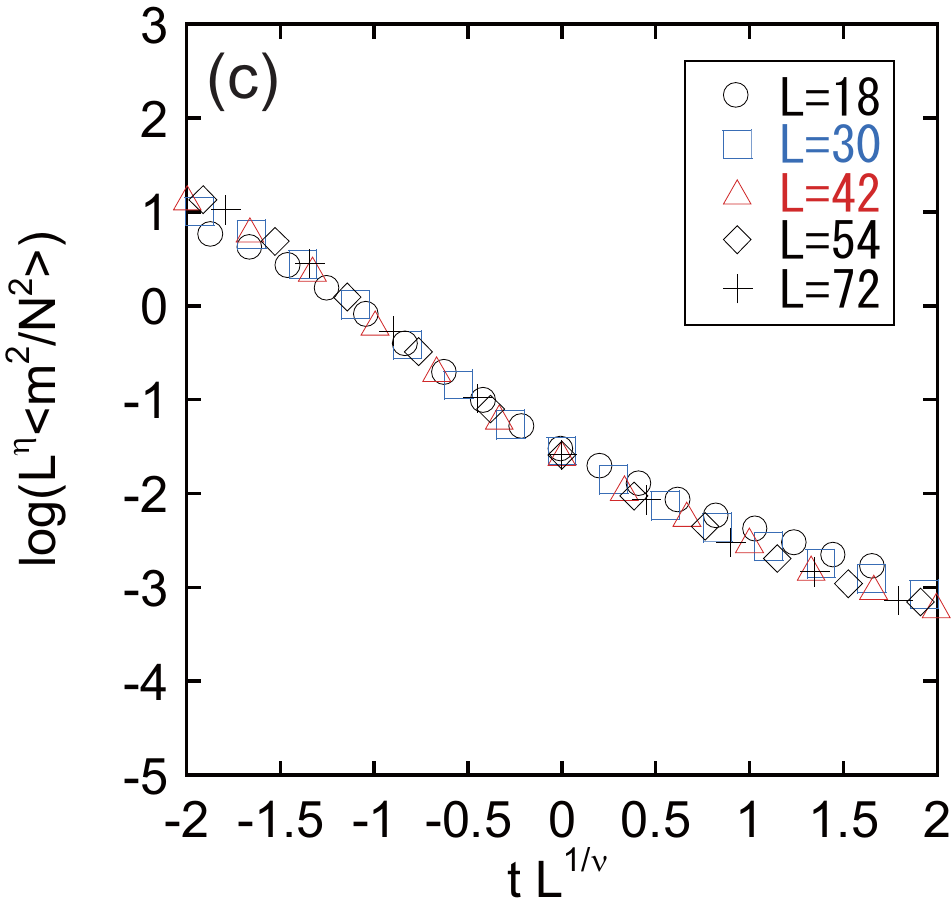}
     \includegraphics[width=62mm]{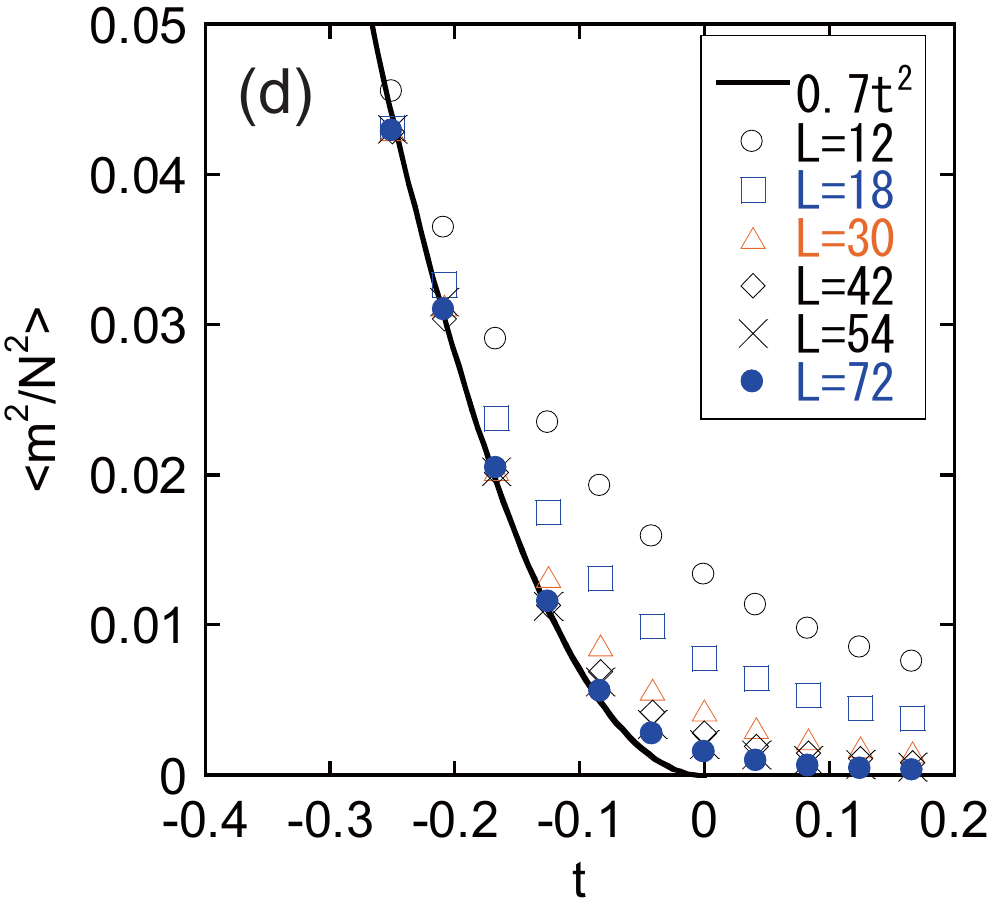}
  \end{center}
\caption{ (color online)  
Critical exponents for the order parameter $m$ in the elastic model (\ref{Ham}). 
(a) $U_4(m)$ is plotted as a function of $t L^{\ 1/\nu}$ for several system sizes $L$. The critical exponent $\nu=1.8$ is adopted for $T_{\rm c2}=0.12$.  
(b) For the order parameter $m$, $a(L,L')$ is given as a function of $T$. 
Symbols denote $(L,L')=(24,12)$ ($+$), $(36,18)$ ($\times$), $(48,24)$ ($\diamond$), $(60,30)$ ($\Box$), $(72,36)$ ($\circ$). 
(c) $\log (L^{\eta}  \langle \frac{m^2}{N^2} \rangle)$ is plotted as a function of $t L^{\ 1/\nu}$ for several system sizes $L$ with the use of $\nu=1.8$ and  $\eta = 1.15$. 
(d) $\langle \frac{m^2}{N^2} \rangle$ is shown as a function of $t$ with several system sizes $L$. 
The solid line denotes the function of $\langle \frac{m^2}{N^2} \rangle=0.7 t^2$. 
 }
\label{FigS1}
\end{figure}

Here we estimate the critical exponents $\nu$, $\eta$, $\beta$ and $\gamma$ for the order parameter $m$ in the elastic model (\ref{Ham}). 
The Binder parameter has a relation between $m$, $t$, $L$, and $\nu$, given by  
\begin{eqnarray}
U_4(m) && =\Psi (t L^{\ 1/\nu}),  
\end{eqnarray}
where $t \equiv \frac{T-T_{\rm c2}}{T_{\rm c2}}$ and $\Psi$ is a scaling function. 
We plot $U_4(m)$ vs. $t L^{\ 1/\nu}$ in Fig.~\ref{FigS1} (a) with $\nu=1.8$. 
From a clear crossing of $U_4(m)$ in Fig.~\ref{Fig_mag} (a), we estimated $T_{\rm c2}=0.12$. 
We find that the data collapse well onto a single curve, and $\nu \simeq 1.8$ is justified. 

Next, we estimate $a(L,L')$ for the order parameter $m$. 
We apply Eq.~(\ref{aL}) with the replacement of $M$ with $m$. 
We give  $a(L,L')$ as a function of $T$ in Fig.~\ref{FigS1} (b). 
We estimate $a(L,L') \simeq 0.85$ at the crossing point ($T_{\rm c2}$). Thus we have $\eta \simeq 1.15$. 

In order to check whether the estimated values for $\nu$ and $\eta$ are valid, 
we also plot $\log (L^{\eta}  \langle \frac{m^2}{N^2} \rangle)$ vs $t L^{\ 1/\nu}$ in Fig.~\ref{FigS1} (c) with $\nu=1.8$ and  $\eta = 1.15$, making use of the relation, 
\begin{eqnarray}
\Big \langle \frac{m^2}{N^2} \Big \rangle && =L^{2-\eta-d} f (t L^{\ 1/\nu}) \\  
 && =L^{-\eta} f (t L^{\ 1/\nu}), \nonumber
\end{eqnarray}
where $f$ is a scaling function. 
We find that the data collapse well onto a single curve, and we conclude that the values of $\nu$ and $\eta$ are valid.

Using the hyperscaling relation for $d=2$, we have 
\begin{eqnarray}
\beta = \frac{\nu (d-2 +\eta)}{2}=\frac{\nu \eta}{2} \simeq 1.0. 
\end{eqnarray}
We find that the value of $\beta$ is about 1 and larger than $1/2$. 
Because $\langle \frac{m^2}{N^2} \rangle = m_{\rm S}^2 + k_{\rm B} T \frac{\chi}{N}$ for $T < T_{\rm c2}$, where $m_{\rm S}=\langle \frac{m}{N} \rangle$ and $\chi=\frac{\langle m^2 \rangle - \langle m \rangle^2   }{N k_{\rm B} T  }$,  
 $\langle \frac{m^2}{N^2} \rangle \propto (t^{\beta})^2$ if $1 \ll L$. 
We plot $\langle \frac{m^2}{N^2} \rangle$ as a function of $t$ with several system sizes $L$ in Fig.~\ref{FigS1} (d). We also give a solid line for $\langle \frac{m^2}{N^2} \rangle = 0.7 t^2$ in Fig.~\ref{FigS1} (d), and 
we find that this line is regarded as the asymptotic line of $m_{\rm S}^2$ for larger $L$. Thus we conclude $\beta \simeq 1$. 
With the use of the hyperscaling relation, $\gamma=\nu (2-\eta)$, $\gamma=1.53$ is derived. 

\subsection{Critical exponents for the long-range interaction model}
\label{sec_}

\begin{figure}[t]
  \begin{center}
     \includegraphics[width=60mm]{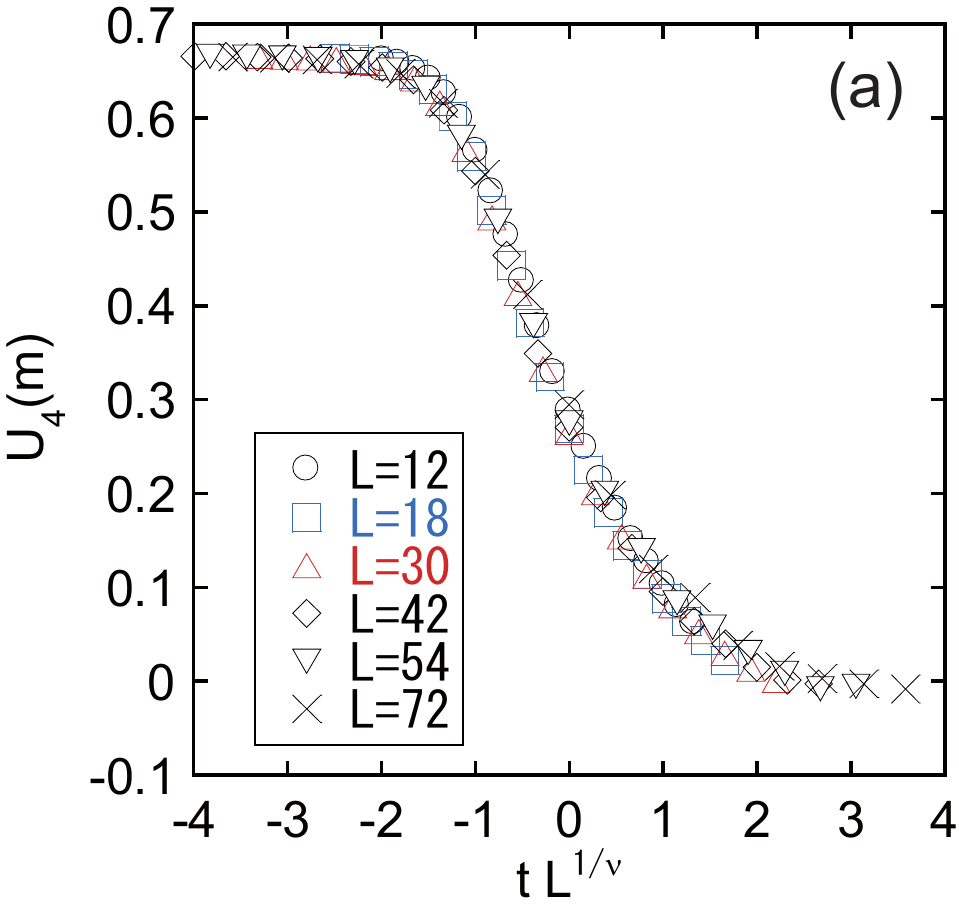}
     \includegraphics[width=62mm]{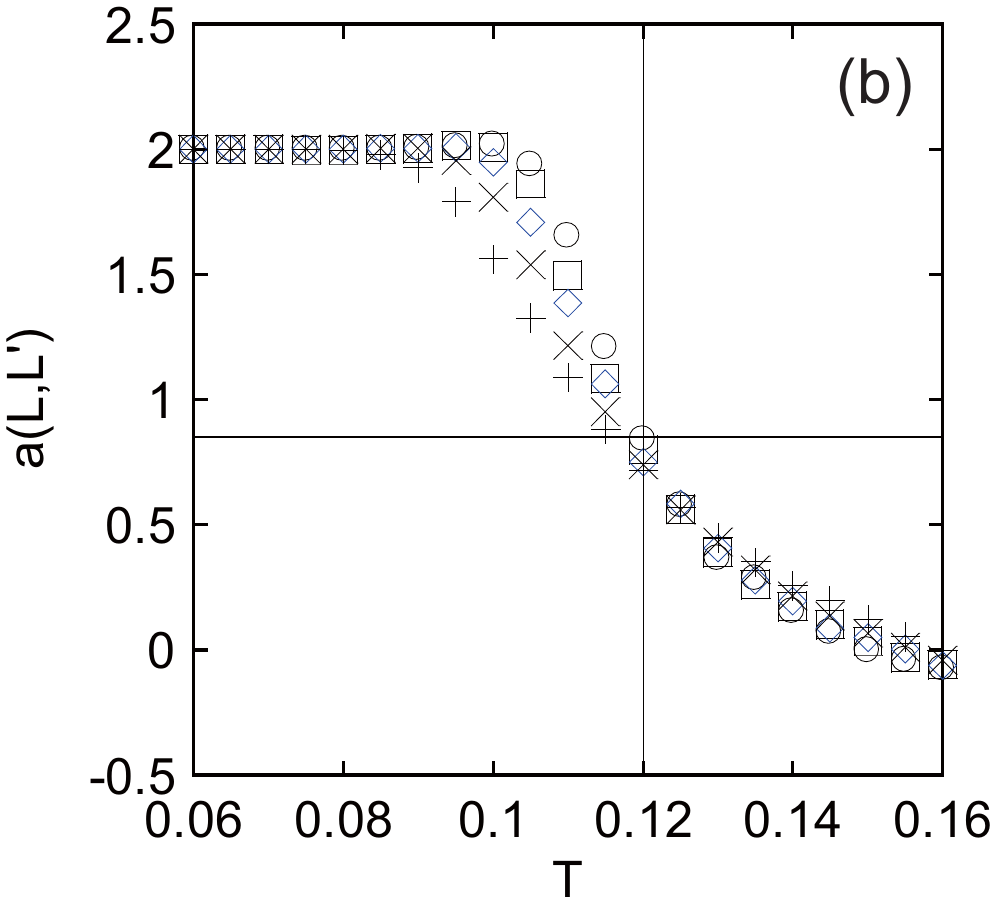}
     \includegraphics[width=60mm]{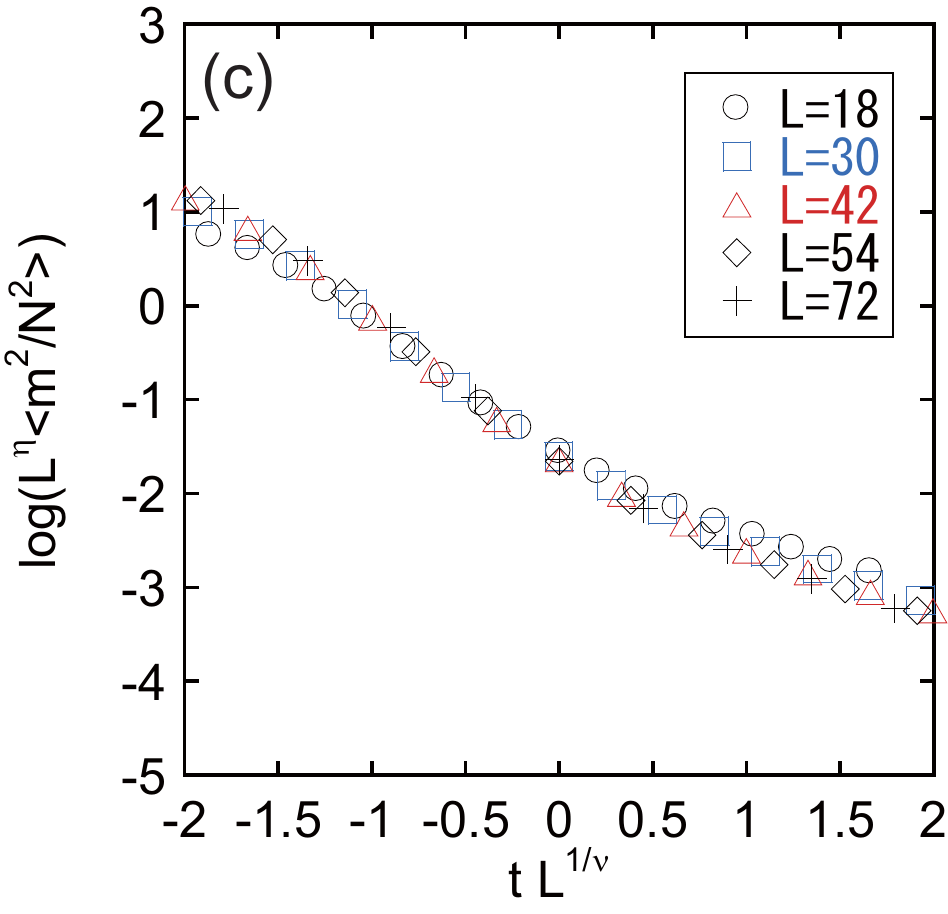}
     \includegraphics[width=62mm]{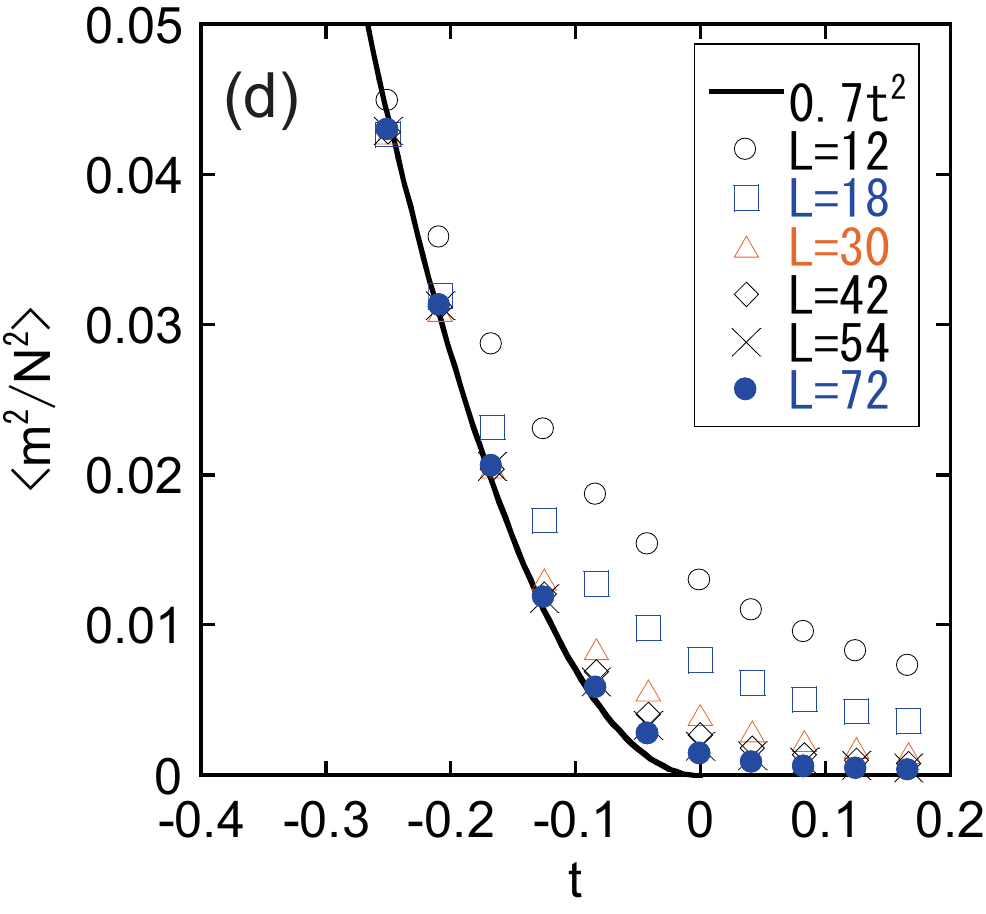}
  \end{center}
\caption{ (color online)  
Critical exponents for the order parameter $m$ in the long-range interaction model (\ref{Ham_prime}). 
(a) $U_4(m)$ is plotted as a function of $t L^{\ 1/\nu}$ for several system sizes $L$. The critical exponent $\nu=1.8$ is adopted for $T_{\rm c2}=0.12$.  
(b) For the order parameter $m$, $a(L,L')$ is given as a function of $T$. 
Symbols denote $(L,L')=(24,12)$ ($+$), $(36,18)$ ($\times$), $(48,24)$ ($\diamond$), $(60,30)$ ($\Box$), $(72,36)$ ($\circ$). 
(c) $\log (L^{\eta}  \langle \frac{m^2}{N^2} \rangle)$ is plotted as a function of $t L^{\ 1/\nu}$ for several system sizes $L$ with the use of $\nu=1.8$ and  $\eta = 1.15$. 
(d) $\langle \frac{m^2}{N^2} \rangle$ is shown as a function of $t$ with several system sizes $L$. 
The solid line denotes the function of $\langle \frac{m^2}{N^2} \rangle=0.7 t^2$. 
 }
\label{FigS2}
\end{figure}

Next we perform the same analyses in Figs~\ref{FigS2} (a)-(d) 
for the critical exponents for $m$ in the long-range interaction model (\ref{Ham_prime}).  Here we find excellent agreements for the scaling properties between the two models, and we have a conclusion that these two models are 
equivalent with respect to the criticality at $T_{\rm c2}$.   

\begin{figure}[h,t]
  \begin{center}
     \includegraphics[width=60mm]{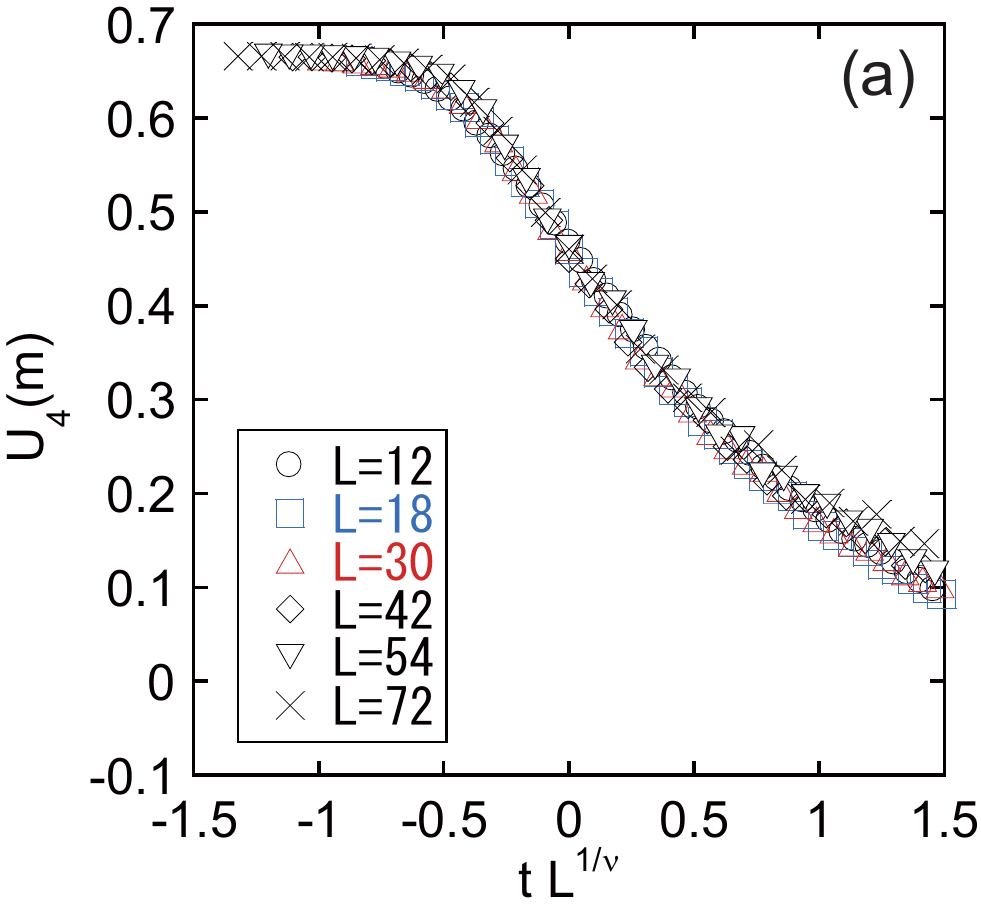}
     \includegraphics[width=62mm]{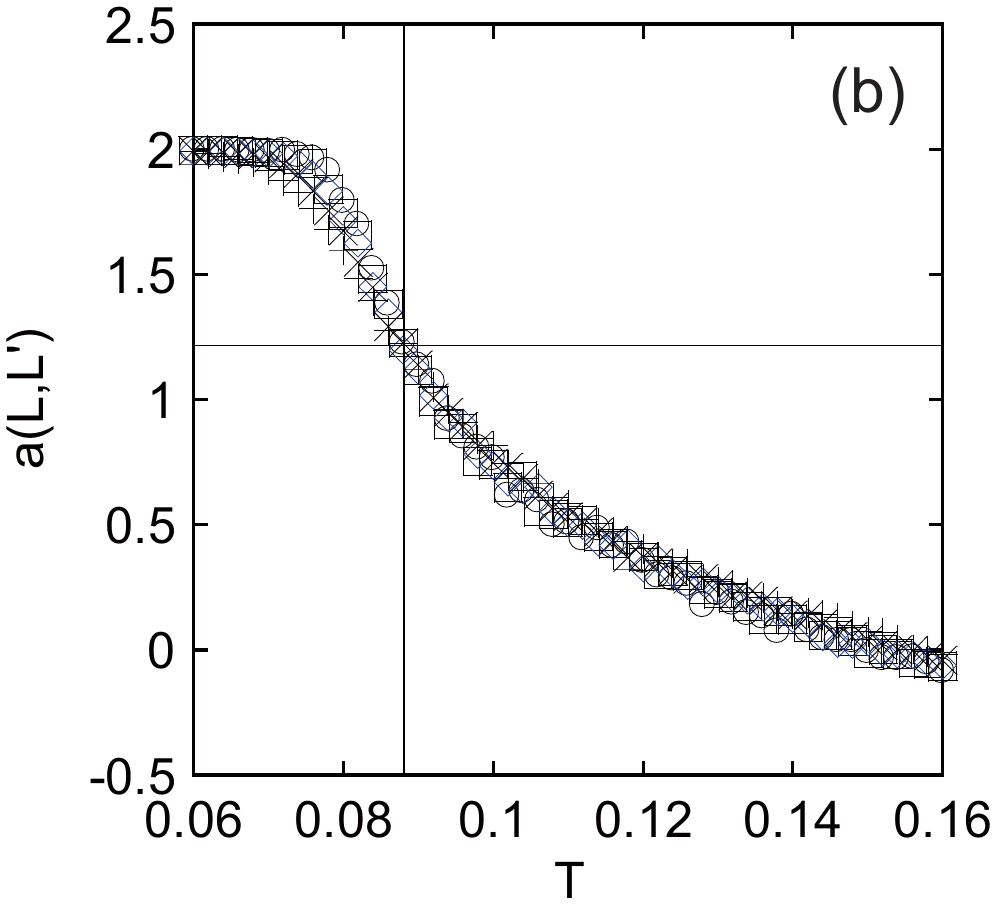}
     \includegraphics[width=60mm]{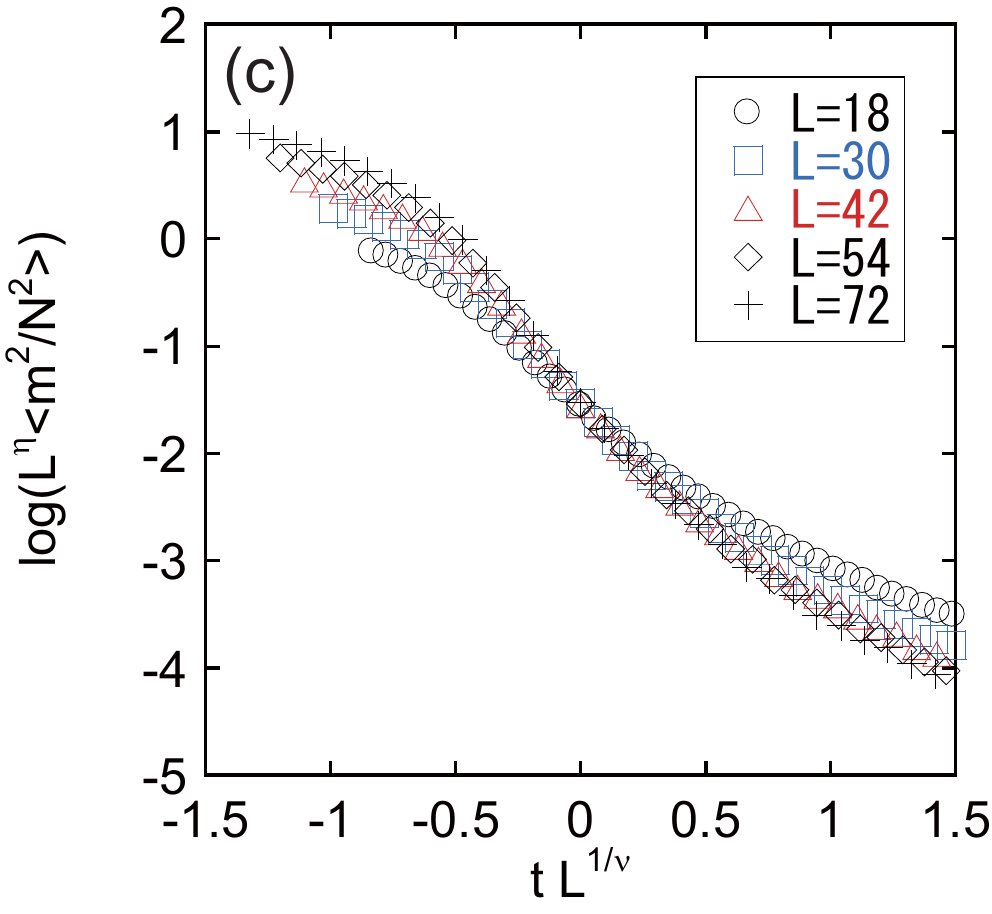}
  \end{center}
\caption{ (color online)  
Critical properties of the TIAFF model (2). 
(a) $U_4(m)$ is plotted as a function of $t L^{\ 1/\nu}$ for several system sizes $L$. The critical exponent $\nu=3.0$ is adopted for $T_{\rm c2}=0.088$.  
(b) For the order parameter $m$, $a(L,L')$ is given as a function of $T$. 
Symbols denote $(L,L')=(24,12)$ ($+$), $(36,18)$ ($\times$), $(48,24)$ ($\diamond$), $(60,30)$ ($\Box$), $(72,36)$ ($\circ$). 
(c) $\log (L^{\eta}  \langle \frac{m^2}{N^2} \rangle)$ is plotted as a function of $t L^{\ 1/\nu}$ for several system sizes $L$ with the use of $\nu=3.0$ and  $\eta = 0.78$. 
 }
\label{FigS3}
\end{figure}

To investigate the difference of the critical properties between the TIAFF model (\ref{Ham_I}) and the equivalent two models (the elastic model and the LRI model), we study the  critical exponents for the TIAFF model (\ref{Ham_I}) in the same way. 
We depict $U_4(m)$ vs. $t L^{\ 1/\nu}$ in Fig.~\ref{FigS3} (a) with $\nu=3.0$. 
From a clear crossing of $U_4(m)$ in Fig.~\ref{Fig_mag} (b), we estimate
$T_{\rm c2}=0.088$. The data collapse well onto a single curve. 
Next we show  $a(L,L')$ for $m$ as a function of $T$ in Fig.~\ref{FigS3} (b). 
At the crossing point ($T_{\rm c2}$), we find $a(L,L') \simeq 1.22$ and thus $\eta \simeq 0.78$. 
Making use of the exponents $\nu=3.0$ and $\eta = 0.78$, 
we plot $\log (L^{\eta}  \langle \frac{m^2}{N^2} \rangle)$ vs $t L^{\ 1/\nu}$ 
in Fig.~\ref{FigS3} (c).  
However, the data do not collapse onto a single curve. 
We consider that this inconsistency in the scaling properties is 
related to the specialty of the BKT point although the relation between  
$M^2$ and $m$ is not trivial. 
The above-mentioned analyses lead to an important conclusion: 
the critical properties of the order parameter $m$ for the elastic model and 
the LRI model are different from those for the TIAFF model (\ref{Ham_I}).

Here we find that the fluctuation of the uniform ferromagnetic-like mode is 
essential for the critical nature at $T_{\rm c2}$, which causes a new class of critical phenomena. 
Furthermore, the difference of the uniform magnetization 
between the present case and the two sublattice ferrimagnetic state of the 
MF theory should be noted. The latter appears when  $J_2/J_1 < -0.4$ (relatively large values of $J_2$) for the three sublattice ferrimagnetic model with the MF theory~\cite{Mekata}.  In this case, each sublattice magnetization for three sublattices gives $\beta=1/2$ and the uniform magnetization $m$ gives $\beta=3/2$, and thus  
$m^2$ shows a convex shape. 
However, the uniform ferromagnetic mode in this case is an irrelevant order parameter. That is, the uniform susceptibility does not diverge at the critical point~\cite{Mekata}, and the value of $U_4(m)$ is almost zero $U_4(m) < O(10^{-2})$~\cite{calc}. 
Thus, this critical process is different from the present one.

\section{Summary}
\label{summary}

In summary, we discovered a new type of critical phenomena in the ordering process of the SC model in the triangular lattice with the elastic interaction. 
When the elastic interaction is relatively weak, the effect of the elastic interaction changes the critical property at the low-temperature endpoint of the BKT phase, where a new class of critical phenomena of the uniform magnetization is realized. We found that the elastic interaction and the ferromagnetic infinite-range interaction are equivalent for the TIAFF model (\ref{Ham_I}) in the region of relatively weak elastic coupling, and thus the same ordering process appears in the model (\ref{Ham_prime}).  
If the elastic interaction is relatively strong, the SC model shows a first-order transition between the disordered and ferrimagnetic-like phases and the equivalence does not hold anymore. 
The BKT phase exists as an intermediate phase between disordered and ferrimagnetic-like phases. Below the higher critical temperature $T_{\rm c1}$ the BKT phase appears, but the uniform magnetization $m$ is not enhanced in the BKT phase, i.e., the symmetry between HS-rich and LS-rich states holds and no significant volume fluctuation appears, where the long-range interaction (the elastic interaction) is negligible. 
On the other hand, in the ferrimagnetic-like phase the uniform magnetization $m$ appears and a large fluctuation of $m$ in space is generated around $T_{\rm c2}$, and thus the long-range interaction (the elastic interaction) plays an important role. 
This provides a new scenario for the phase transitions at the two endpoints of the intermediate BKT phase.

\section*{Acknowledgments}
The present work was supported by 
Grants-in-Aid for Scientific Research C (No.26400324, 25400391) from MEXT of Japan. The authors thank the Supercomputer Center, the Institute for Solid
State Physics, the University of Tokyo for the use of the
facilities.


\begin{thebibliography}{99}
\bibitem{calc}
We estimated by a numerical integration $U(m)_4$ at the N\'eel temperature of the three-sublattice MF model for $J_2/J_1 < -0.4$.
\end{thebibliography}
\end{document}